\documentclass[12pt]{iopart}

\usepackage{iopams}  

\usepackage{graphicx}
\usepackage{epsfig} 
\usepackage{dcolumn}
\usepackage{bm}
\usepackage{epstopdf}
\usepackage{multirow}
\usepackage{amsmath}
\usepackage{booktabs}
\usepackage{color}
\usepackage{gensymb}
\usepackage{xfrac}
\usepackage{kantlipsum}  
\usepackage{hyperref}
\usepackage{ulem}
\usepackage{braket}
\usepackage{physics}
\usepackage[dvipsnames]{xcolor}

\usepackage[T1]{fontenc}
\usepackage{mathptmx}

\newcommand{\cS}{$c \, ^3\Sigma^+_1 \ $}
\newcommand{\cSone}{$c \, ^3\Sigma^+_1 \ $}

\newcommand{\aS}{$a \, ^3\Sigma^+ \ $}
\newcommand{\XS}{$X \, ^1\Sigma^+ \ $}
\newcommand{\BP}{$B \, ^1\Pi \ $}
\newcommand{\bP}{$b \, ^3\Pi_1 \ $}
\begin{document}

\title{Efficient Pathway to NaCs Ground State Molecules}

\author{Claire Warner, Niccol\`{o} Bigagli, Aden Z. Lam, Weijun Yuan, \mbox{Siwei Zhang}, Ian Stevenson, and Sebastian Will}

\address{Department of Physics, Columbia University, New York, New York 10027, USA}
\date{\today}

\begin{abstract}
We present a study of two-photon pathways for the transfer of NaCs molecules to their rovibrational ground state. Starting from NaCs Feshbach molecules, we perform bound-bound excited state spectroscopy in the wavelength range from 900~nm to 940~nm, covering more than 30 vibrational states of the $c \, ^3\Sigma^+$, $b \, ^3\Pi$, and $B \, ^1\Pi$ electronic states. Analyzing the rotational substructure, we identify the highly mixed \cSone$\ket{v=22}$ $\sim$ \bP$\ket{v=54}$ state as an efficient bridge for stimulated Raman adiabatic passage (STIRAP). We demonstrate transfer into the NaCs ground state with an efficiency of up to 88(4)\%. Highly efficient transfer is critical for the realization of many-body quantum phases of strongly dipolar NaCs molecules and high fidelity detection of single molecules, for example, in spin physics experiments in optical lattices and quantum information experiments in optical tweezer arrays.
\end{abstract}

\maketitle

\section{Introduction}

Ultracold molecules~\cite{krems2008cold, carr2009cold, moses2017new, bohn2017cold} are an emerging platform for the investigation of new frontiers in many-body quantum physics~\cite{micheli2006toolbox, lahaye2009physics, baranov2012condensed, yan2013observation, altman2021quantum, christakis2023probing},  quantum information~\cite{demille2002quantum, park2017second, sawant2020ultracold, gregory2021robust, holland2022demand, bao2022dipolar}, and quantum chemistry~\cite{krems2008cold, ospelkaus2010controlling, quemener2012ultracold, ye2018collisions, hu2019direct, gregory2020loss, liu2021precision, bause2023ultracold, park2023feshbach}. Many of the proposed applications require large samples of ultracold molecules at high phase-space densities. The highest phase-space densities in molecular systems have so far been achieved by assembling ultracold heteronuclear molecules from ultracold atoms. First, loosely bound molecules are associated in the vicinity of a Feshbach resonance~\cite{chin2010feshbach}. Then, the molecules are transferred to the rovibrational ground state using a coherent two-photon process that utilizes an electronically excited  state as a bridge between the Feshbach and the ground state. The workhorse technique for ground state transfer of bialkali molecules is stimulated Raman adiabatic passage (STIRAP)~\cite{vitanov2017stimulated}. This approach has been demonstrated for homonuclear molecules~\cite{lang2008ultracold, danzl2008quantum, leung2021ultracold} and heteronuclear molecules~\cite{ni2008high, takekoshi2014ultracold, molony2014creation, park2015ultracold, guo2016creation, rvachov2017long, seesselberg2018modeling, liu2019observation, voges2020ultracold, cairncross2021assembly, stevenson2022ultracold}. The highest efficiencies for STIRAP transfer reported so far are around 90\%, for RbCs~\cite{molony2016measurement} and NaRb~\cite{christakis2023probing,guo2018dipolar}. Often, the key to optimizing transfer efficiencies is a careful survey of the excited state structure, and ultimately the selection of an ideal intermediate state~\cite{park2015two, molony2016production, guo2017high}.

High STIRAP efficiency minimizes the loss in phase-space density when an ensemble of Feshbach molecules is transferred to the ground state. This has been key in the recent creation of degenerate Fermi gases of ground state molecules~\cite{de2019degenerate, duda2023transition} and will be crucial for achieving Bose-Einstein condensation of molecules. Additionally, molecules are typically detected via reversal of the assembly process, followed by imaging of the constituent atoms. The efficiency of reverse STIRAP directly impacts the detection fidelity. High detection fidelity will be critical for the preparation and characterization of novel many-body phases of dipolar molecules, such as crystalline bulk phases~\cite{buchler2007strongly}, exotic density order~\cite{capogrosso2010quantum, baranov2012condensed, schmidt2022self}, or spin order in optical lattices~\cite{micheli2006toolbox, gorshkov2011tunable}. 

Recently, the assembly of NaCs ground state molecules has been reported in optical tweezers~\cite{cairncross2021assembly} and a bulk system~\cite{stevenson2022ultracold}. NaCs is of particular interest due to its large dipole moment of 4.75(20) Debye~\cite{dagdigian1972molecular} in the rovibrational ground state, which promises access to novel strongly correlated and highly entangled quantum phases~\cite{lahaye2009physics, baranov2012condensed}. This large dipole moment also makes NaCs an ideal candidate for microwave shielding~\cite{karman2018microwave, lassabliere2018controlling, anderegg2021observation, schindewolf2022evaporation} and resonant shielding~\cite{avdeenkov2006suppression, martinez2017adimensional, matsuda2020resonant, li2021tuning}, two methods to suppress losses from inelastic two-body collisions. So far, finding a versatile, efficient pathway to the ground state for NaCs has been a challenge. Raman-Rabi transfer with an efficiency of 82(10)\% via the \cSone$\ket{v=26}$ excited state has been reported in Ref.~\cite{cairncross2021assembly} for ground state transfer of a single NaCs molecule in an optical tweezer. However, this scheme is challenging for spatially extended systems, such as large molecular tweezer arrays and bulk gases, as it requires highly uniform laser intensity for precise $\pi$-pulses. STIRAP transfer is less sensitive to experimental imperfections, but in earlier work by our group \cite{stevenson2022ultracold} it was limited to an efficiency of 55(3)\%. 

In this work, we report on the preparation of spin-polarized gases of NaCs ground state molecules with a transfer efficiency of up to 88(4)\%. This improvement stems from an in-depth analysis of the electronically excited states of NaCs. We find \cSone$\ket{v=22}$ to have favorable properties and use it as an intermediate state for STIRAP transfer into two distinct hyperfine states. This advance is an important step towards the preparation of high-phase space density NaCs gases and the high-fidelity detection of single molecules in tweezer arrays or extended bulk samples via quantum gas microscopy.

\section{Approach} 

\begin{figure} 
    \centering
    \includegraphics[width = 8.6 cm]{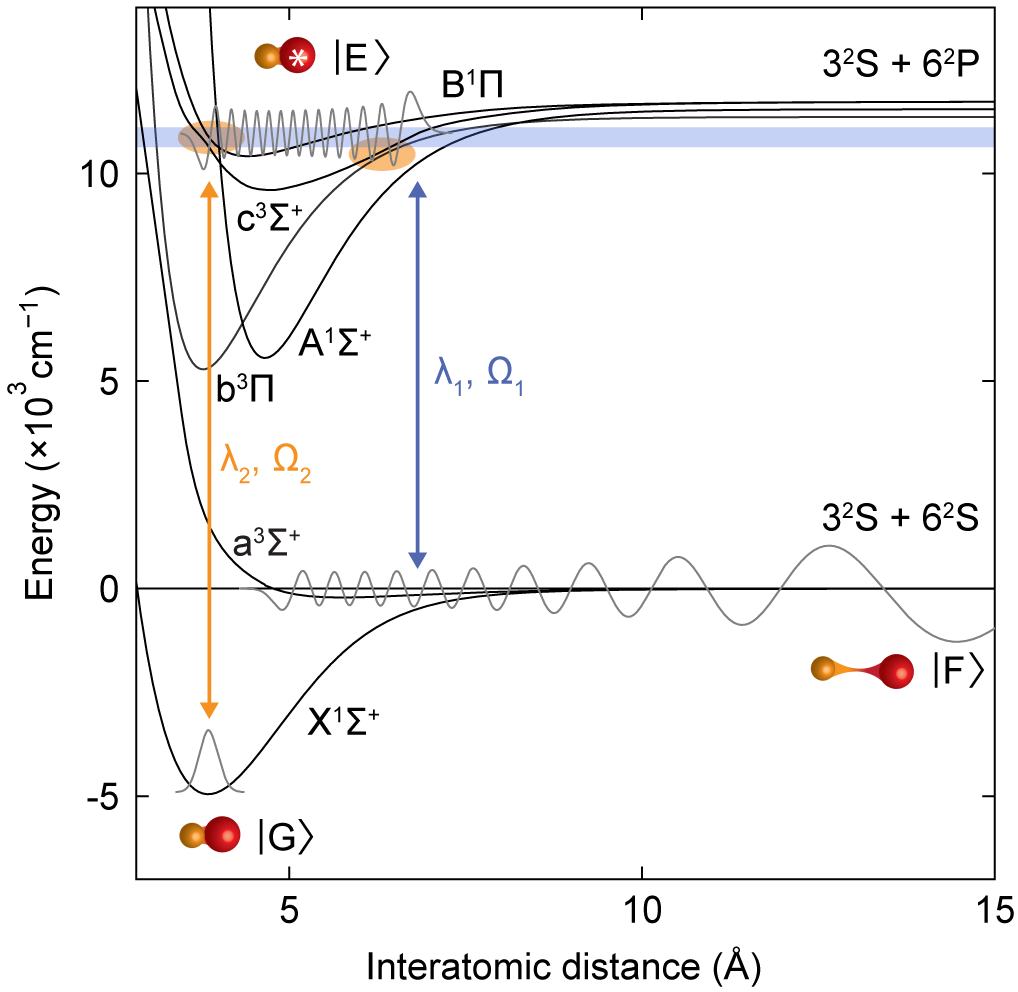}
    \caption{Potential energy curves of the NaCs molecule below the Na $3^2S$ + Cs $6^2P$ dissociation limit. States are labeled following Hund's case (a). The vibrational wavefunctions of the Feshbach $\ket{F}$ and ground $\ket{G}$ states are plotted, as well as the $\ket{E} = \ $\cSone$\ket{v=22}$ excited state. The wavelengths and Rabi frequencies of the STIRAP up-leg (down-leg)  transitions are labeled $\lambda_{1(2)}$ and $\Omega_{1(2)}$, respectively. The blue-shaded region indicates the range of $\lambda_1$ studied in Figure~\ref{fig:3_full_spectrum}. The orange-shaded ovals indicate regions of avoided crossings in the potentials, which lead to significant shifts of vibrational state locations. 
    }
    \label{fig:1_ELD}
\end{figure}

STIRAP~\cite{vitanov2017stimulated} is a scheme for coherent population transfer between the low energy states $\ket{F}$ and $\ket{G}$ of a $\Lambda$-type system via an intermediate excited state $\ket{E}$. Here, $\ket{F}$ denotes the Feshbach molecular state, $\ket{G}$ the rovibrational ground state, and $\ket{E}$ an electronically excited molecular state. $\ket{F}$ and $\ket{G}$ are coupled to $\ket{E}$ by two phase-coherent lasers, as illustrated in Figure~\ref{fig:1_ELD}. For the up-leg laser, which couples $\ket{F}$ and $\ket{E}$, the Rabi frequency is denoted by $\Omega_1$ and the wavelength by $\lambda_1$. For the down-leg laser, which couples $\ket{G}$ and $\ket{E}$, the Rabi frequency is denoted by $\Omega_2$ and the wavelength by $\lambda_2$. As a result of the coupling, a dark state $\cos\theta(t)  \ket{F} - \sin \theta(t) \ket{G}$ emerges, where $\theta(t) = \tan^{-1} (\Omega_1/\Omega_2)$. This dark state can be continuously transformed from $\ket{F}$ to $\ket{G}$ by adiabatically varying $\Omega_1$ and $\Omega_2$.

The objective of this work is to identify a state $\ket{E}$ that optimizes STIRAP efficiency for NaCs. Despite the fact that $\ket{E}$ does not contribute to the dark state, its properties strongly impact the transfer efficiency. We demonstrate that the choice of a suitable intermediate state can significantly enhance STIRAP transfer efficiency. Our search is guided by the following three criteria: (1) $\ket{E}$ should feature strong coupling to both $\ket{F}$ and $\ket{G}$. The strength of this coupling is determined by Franck-Condon factors (FCFs), transition dipole moments, and selection rules. (2) The Rabi coupling $\Omega_{1(2)}$ should be on the order of, or larger than, the excited state linewidth $\Gamma$. (3) The excited state should be well-isolated, such that the relevant states resemble a three-level system as much as possible. 

We start by analyzing the electronic potential energy curves for the ground~\cite{docenko2006coupling} and excited~\cite{docenko2006high, zaharova2006experimental, zaharova2007b, grochola2011spin} electronic states of NaCs (see Figure~\ref{fig:1_ELD}). The initial Feshbach state is $\ket{F} = a^3\Sigma^+ \ket{v = 23}$  and the vibrational ground state is $\ket{G} = X^1\Sigma^+ \ket{v = 0}$. The large spin-orbit coupling of Cs leads to avoided crossings between the \cS potential and the \BP and \bP potentials, as indicated in Figure~\ref{fig:1_ELD}. As a result, the nominally spin-forbidden \aS $\leftrightarrow$ \BP and \XS $\leftrightarrow$ \cSone transitions are allowed~\cite{zaharova2007b, grochola2011spin, zabawa2011formation}, such that many of the \cS and \BP vibrational states are expected to have mixed spin character\footnote{NaCs follows Hund's case (c) rather than (a).}. The spin composition of $\ket{F}$, shown in Table~\ref{tab:1_feshbach_molecules}, determines which intermediate spin states $\ket{E}$ couple to $\ket{F}$ and ultimately to $\ket{G}$ based on selection rules for dipole transitions. The Feshbach state is dominantly composed of spin states with triplet character, such that coupling to the singlet ground state benefits from an excited state with mixed spin character, as expected for the \cSone $\sim$ \BP complex. We therefore focus our investigation on this complex. Conveniently, the up-leg and down-leg transition wavelengths are in the range of standard lasers.

\begin{table}
    \centering
    \caption{Decomposition of the Feshbach molecule wavefunction ($m_F = m_S + m_I = 4$) in the $\ket{ S, \ m_S; \ m_{I_{\mathrm{Na}}}, \ m_{I_{\mathrm{Cs}}} } $ basis, where $S$ is the total electronic spin, $m_S$ is the projection of $S$ on the quantization axis, and $m_{I_{\mathrm{Na(Cs)}}}$ is the nuclear spin projection of Na (Cs) on the quantization axis. $m_I = m_{I_{\mathrm{Na}}} + m_{I_{\mathrm{Cs}}}$ is the total nuclear spin projection.}
    \begin{tabular}{l  c  c  c } 
    \hline \hline
    $\ket{ S, \ m_S; \ m_{I_{\mathrm{Na}}}, \ m_{I_{\mathrm{Cs}}} } $ & $m_I$ &  Contribution  & Channel \\ \hline 
    $\ket{ 1, \ -1;\ 3/2,\ 7/2 } $ & 5 & 19.7\%  & Open\\
    $\ket{ 0,\ 0;\  3/2,\ 5/2 } $ & 4 & 25.5\% & Closed\\
    $\ket{ 1,\ 0;\ 3/2,\ 5/2 } $ & 4 & 16.8\% & Closed\\
    $\ket{ 0,\ 0;\ 1/2,\ 7/2 } $ & 4 & 1\% & Closed\\
    $\ket{ 1, \ 1; \ 3/2,\  3/2 } $ & 3 & 31.1\% & Closed \\
    $\ket{ 1,\ 1;\  1/2,\ 5/2 } $ & 3 & 4.9\% & Closed\\
    $\ket{ 1,\ 1; -1/2,\ 7/2 } $ & 3 & 1\% & Closed\\
    \hline \hline
    \end{tabular}
    \label{tab:1_feshbach_molecules}
\end{table}

\begin{figure} 
    \centering
    \includegraphics[width = 3.39 in]{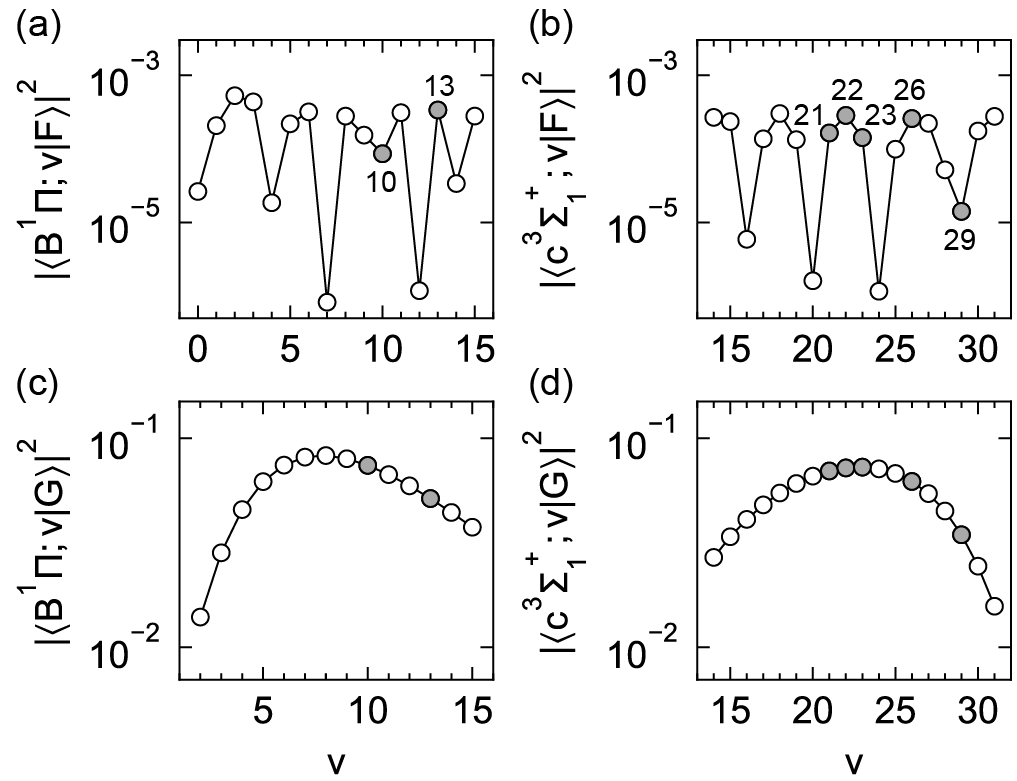}
    \caption{Relevant Franck-Condon Factors for a STIRAP pathway involving the \cSone $\sim$ \BP complex. The vibrational wavefunctions have been calculated using a Numerov-Cooley solver (LEVEL16)~\cite{le2017level}. FCFs between (a) the Feshbach state $\ket{F} = a^3\Sigma^+ \ket{v = 23}$ and \BP vibrational states, (b) the Feshbach state and the \cS vibrational states, (c) the \BP vibrational states and the ground state $\ket{G} = X^1\Sigma^+ \ket{v = 0}$, and (d) \cS vibrational states and the ground state. Filled points, labeled by their vibrational number $v$, indicate vibrational states that are observed spectroscopically in this work. 
    }
    \label{fig:2_FCFs}
\end{figure}

Using vibrational wavefunctions determined from the known potential energy curves, we calculate the FCFs of the vibrational states in the \cSone $\sim$ \BP complex with both $\ket{F}$ and $\ket{G}$. The results are shown in Figure~\ref{fig:2_FCFs}. The FCFs are relatively constant for the down-leg but vary over several orders of magnitude for the up-leg, depending on the excited vibrational state. Therefore, identifying an intermediate state with strong coupling to the Feshbach state is prioritized. Based on the calculated FCFs, there are several candidate states, including \BP$\ket{v=13}$ and \cSone$\ket{v=22, \ 26}$. However, FCFs alone only give rough guidance as to which states may be suitable for achieving efficient STIRAP transfer. In the following, we carry out a detailed spectroscopic investigation of the properties of these and other potential intermediate states.

\section{One-photon spectroscopy}

We perform bound-bound one-photon spectroscopy on ultracold ensembles of NaCs Feshbach molecules. The method is as follows: in a broad scan, we identify possible up-leg transitions. Then, we perform higher-resolution scans on each feature found in the broad scan, allowing the identification of states that belong to the \cSone $\sim$ \BP complex based on their substructure. Finally, we perform hyperfine-resolved scans to identify specific features for STIRAP transfer\footnote{Our wavemeter has a precision of 1~MHz and an accuracy of 100~MHz.}.

Our experiment begins with a sample of $2.0(4) \times 10^4$ NaCs Feshbach molecules at 350(50)~nK, associated via a magnetic field ramp across the interspecies Feshbach resonance at 864.12(5)~G~\cite{warner2021overlapping, lam2022high}. The molecules are held in a crossed optical dipole trap with trap frequencies $\omega = 2 \pi \times \{30(2), 60(4), 160(10)\}$~Hz at a magnetic field of 863.85(1)~G. The magnetic field is oriented vertically, defining the quantization axis. The up-leg and down-leg lasers co-propagate in the horizontal plane. Laser light linearly polarized parallel (perpendicular) to the quantization axis gives access to $\pi \ (\sigma^{+} / \sigma^-)$ transitions. After exposure, the Feshbach molecules are dissociated into Na and Cs atoms using a reverse magnetic field ramp, and the Cs atoms are imaged at low magnetic field. If the up-leg laser is resonant with an excited state, Feshbach molecules are lost from the sample, which manifests as a reduction of the final Cs atom number. For high-resolution scans and STIRAP transfer, we stabilize the lasers to a high-finesse optical cavity ($\mathfrak{F} \sim 20,000$), narrowing their linewidths to less than 1~kHz. For one-photon spectroscopy, we expose the Feshbach molecules to the up-leg laser, a focused beam ($1/e^2$-radius of 47(5)~$\mu$m) generated by a Ti:Sapphire laser. The exposure times range from 1~ms for coarse to 5~$\mu$s for the highest resolution scans.

\begin{figure*} [t]
    \centering
    \includegraphics[width = \textwidth]{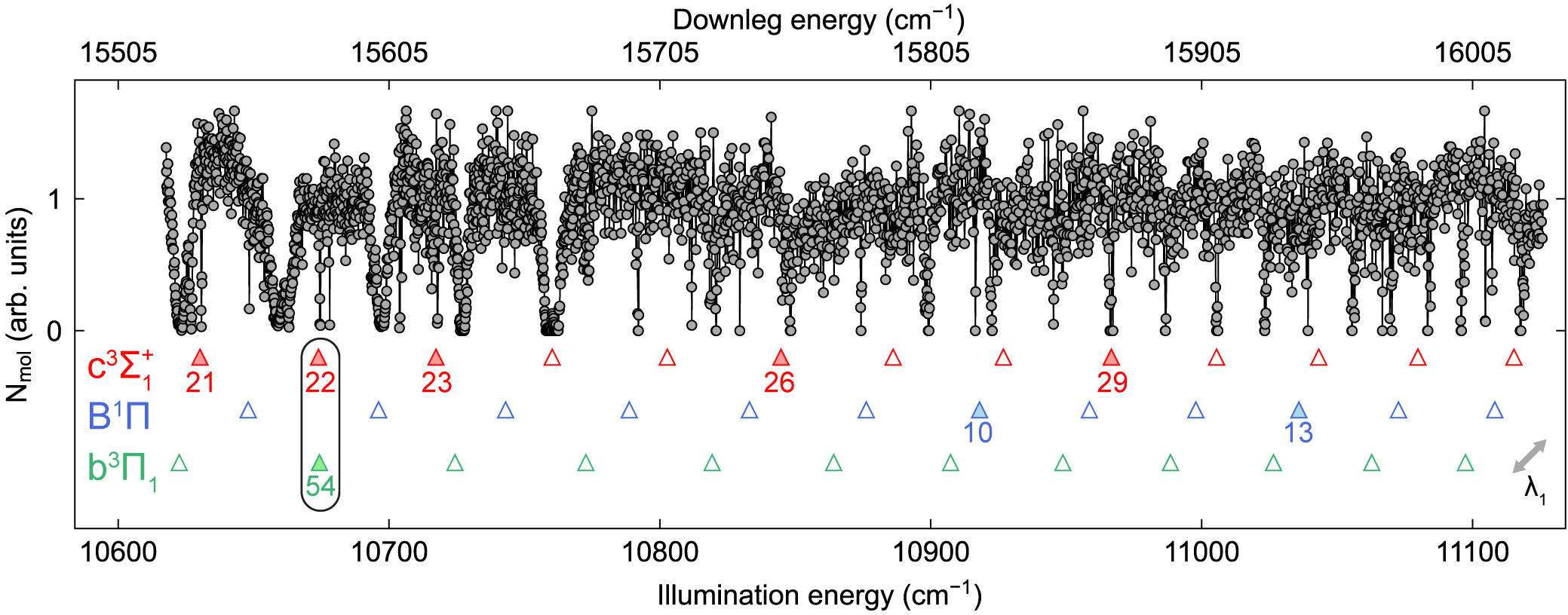}
    \caption{Single-photon spectrum of the electronically excited vibrational states of NaCs. Data points are taken in steps of about 3~GHz, using the up-leg laser with 50~mW of power and 1~ms exposure time. The up-leg laser is linearly polarized at 45 degrees to the quantization axis (illustrated by the grey diagonal arrow in the lower right corner), ensuring that both $\sigma^{\pm}$ and $\pi$ transitions can be accessed. Triangles indicate calculated vibrational state locations~\cite{le2017level} for \cSone(red), \BP (blue), and \bP (green) states from LEVEL16. Filled triangles indicate levels that are uniquely identified based on their rotational substructures and comparison to existing data. They are labeled with the vibrational quantum number $v$. The circled \cSone$\ket{v=22}$ and \bP$\ket{v=54}$ states are overlapping. 
    }
    \label{fig:3_full_spectrum}
\end{figure*}

To coarsely locate vibrational levels, we perform a broad spectroscopic scan, as shown in Figure~\ref{fig:3_full_spectrum}. Loss features correspond to vibrational states in the electronically excited manifolds $B^1\Pi,\ $ $c^3\Sigma^+,\ $ $A^1\Sigma^+,\ $ and $b^3\Pi$, as shown in Figure~\ref{fig:1_ELD}. Overall, we observe 38 vibrational states. For some of the features, the observed locations deviate from calculated vibrational energy levels. This is expected due to avoided crossing of the potential curves and state mixing in this area of the excitation spectrum (see Figure~\ref{fig:1_ELD}). It is interesting to note that some features, located at approximately 10620, 10660, 10700, 10730, and 10760~cm$^{-1}$, show extremely large linewidths. 

\begin{table} 
    \centering
    \caption{ Properties of the observed $\Omega = 1$ lines. $\bar{\Gamma}$ is the averaged linewidth of the $J=1$ features in the hyperfine-resolved scans of each vibrational level.}
    \begin{tabular}{c  c  c  c  c} 
    \hline \hline
     Energy (THz) & Energy (cm$^{-1}$) &  Hund's (a) label & $\bar{\Gamma} / (2 \pi)$ (MHz)   \\ \hline 
    318.695 & 10630.5 & \cS $\ket{v=21}$ & 30(5) \\
    320.007 & 10674.3 & \bP $\ket{v=54}$ & 4(2)  \\
    320.008 & 10674.3 & \cS $\ket{v=22}$ & 6(1)  \\
    321.309 & 10717.7 & \cS $\ket{v=23}$ & 12(3)  \\
    325.130 & 10845.2 & \cS $\ket{v=26}$ & 51(6) \\ 
    327.267 & 10916.5 & \BP $\ket{v=10}$ & 15(2)  \\ 
    328.790 & 10967.3 & \cS $\ket{v=29}$ & 180(50) \\
    330.767 & 11033.2 & \BP $\ket{v=13}$ & 10(2) \\ 
    \hline \hline
    \end{tabular}
    \label{tab:2_line_assignments}
\end{table}

In the next step, we identify features with $\Omega = 1$ substructure, as those features belong to the \cSone $\sim \ $ \BP complex. Here, $\Omega = \Sigma + \Lambda$, where $\Sigma$  and $\Lambda$ are the projections of the total electronic spin and orbital angular momenta, respectively, on the internuclear axis. The value of $\Omega$ for a vibrational state is directly reflected in its rotational structure. States with $\Omega = 0^+$ have $J = 1$ features, where $J = \Omega + R$ and $R$ is the rotational angular momentum of the nuclei. States with $\Omega = 0^-$ have features for $J=0$ and 2. States with $\Omega = 2$ have features for $J=2$. States with $\Omega = 1$ have $J=1$ and 2 features~\cite{herzberg1945molecular}. We perform medium-resolution scans that resolve the rotational substructure for all vibrational features observed in the broad scan to identify $\Omega = 1$ states. Combining the measured rotational substructure with existing NaCs spectroscopic data~\cite{zabawa2012production}, we uniquely identify several of these features as vibrational levels in the \cSone or \BP manifolds. The state assignments are shown in Figure~\ref{fig:3_full_spectrum}. Table~\ref{tab:2_line_assignments} lists these states together with fitted linewidths of the corresponding $J = 1$ features, which are relevant for coupling into the rovibrational ground state. 

\begin{figure*} 
    \centering
    \includegraphics[width = \textwidth]{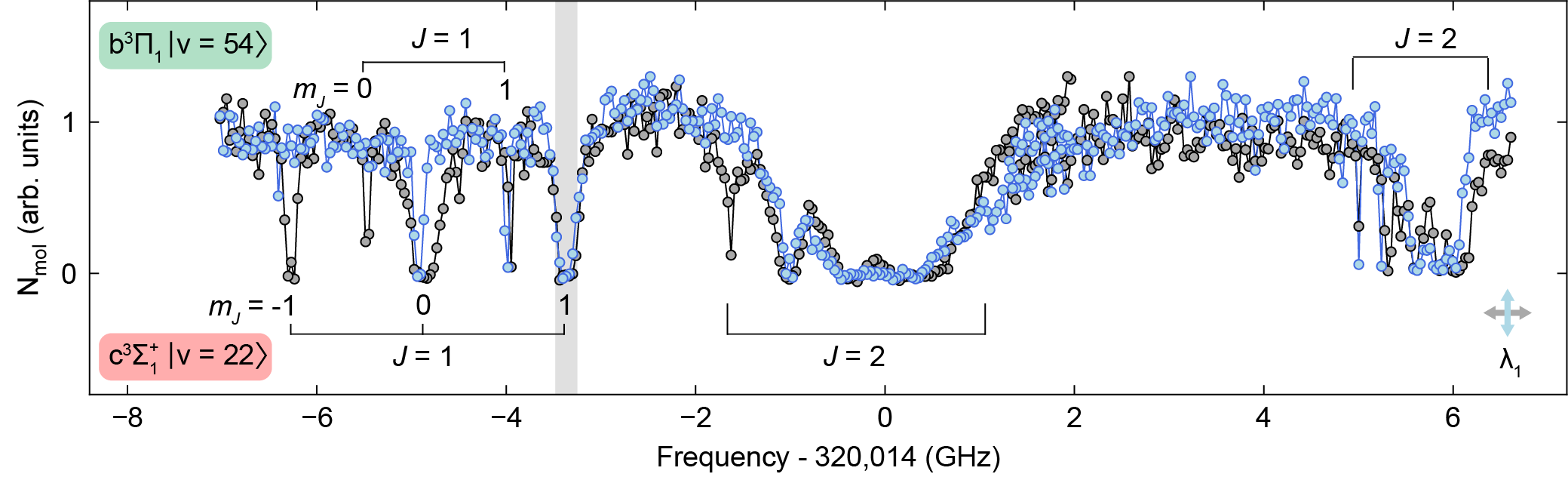}
    \caption{Measured rotational structure of the mixed \cSone$\ket{v=22} \sim$ \bP$\ket{v=54}$ states. Data points are taken in steps of about 30~MHz, using the up-leg laser with 20~mW of power and 200~$\mu$s exposure time. Top (bottom) labels indicate \bP$\ket{v=54}$ (\cS$\ket{v=22}$) rotational states. Blue (grey) data is taken with vertical (horizontal) laser polarization. Grey shading indicates the feature used for STIRAP transfer. The $J=1, \ m_J = -1$ line for \bP$\ket{v=54}$ is not observed. In the horizontally polarized spectrum, we observe all three $m_J$ states for $J=1$. In the vertically polarized spectrum, the $m_J = -1$ line is weakened. We observe a polarization-dependent shift in the $m_J$ lines as a result of the underlying hyperfine structure. }
    \label{fig:4_c22_medium}
\end{figure*}

We identify \cSone$\ket{v=22}$ as a promising candidate state based on its FCF and the relatively narrow linewidth. The rotational state-resolved scan is shown in Figure~\ref{fig:4_c22_medium}. Interestingly, the scan reveals that \cSone$\ket{v=22}$ overlaps with the \bP$\ket{v=54}$ state, with rotational sublevels being separated by less than 1~GHz. In general, the \bP states are not expected to be visible in bound-bound spectroscopy from the Feshbach state due to their extremely weak FCFs. In this case, however, the close proximity of \bP$\ket{v=54}$ and \cSone$\ket{v=22}$ leads to state mixing, which broadens the linewidths of \bP$\ket{v=54}$ features, making them visible in the scan, while narrowing the linewidths of \cSone$\ket{v=22}$ features. 

\begin{figure}
    \centering
    \includegraphics[width = \textwidth]{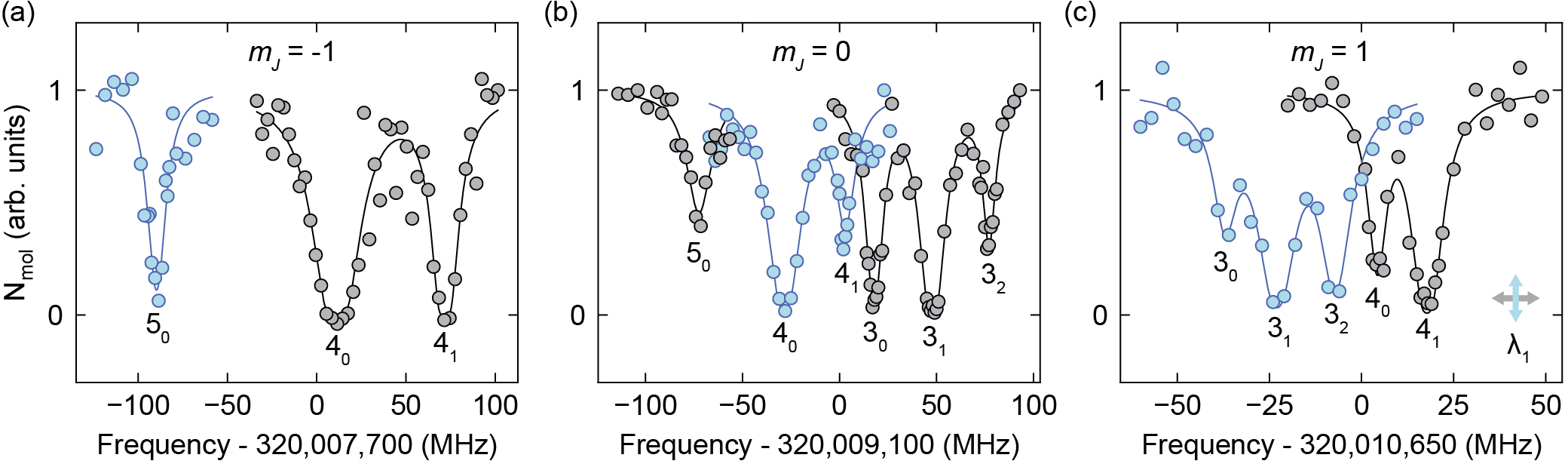}
    \caption{Hyperfine-resolved scan of \cSone$\ket{v = 22, \ J = 1}$ for the (a) $m_J = -1$, (b) $m_J = 0$, and (c) $m_J = 1$ sublevels (observed in Figure~\ref{fig:4_c22_medium}).  Each feature is labeled with $m_{I, n}$. Blue (grey) data points represent data taken with vertical (horizontal) light polarization, represented by the arrows in the lower right corner. Solid lines show fits to eq.~\ref{eq:fit_eqn}; all fit results are shown in Table~\ref{tab:3_fine_spectroscopy}. The light intensities and exposure times are varied to provide sufficient signal strength for fitting.  
    }
    \label{fig:5_c22_narrow}
\end{figure}

As a result, shown in Figure~\ref{fig:5_c22_narrow}, it is possible to resolve the hyperfine structure of the $J = 1, \ m_J = -1, \ 0, \ 1$ sublevels of \cSone$\ket{v=22}$. This is the only excited vibrational state in our scan range for which we observed fully resolved hyperfine structure. We use the structure of the hyperfine features and their dependence on laser polarization to assign the $m_I$ quantum number, labelling them by $m_{I, n}$, where $n$ is a counting index, following a similar convention as Ref.~\cite{gregory2021molecule}. Here, $m_I$ is the projection of $I$ onto the quantization axis and $I = I_{\rm Na } + I_{\rm Cs}$ is the sum of the nuclear spins of Na and Cs. The integer $n$ counts the possible combinations of $m_{I_{\rm Na}}$ and $m_{I_{\rm Cs}}$ that result in the given value of $m_I$ in order of ascending energy. We fit each $m_{I, n}$ feature to an exponential decay model
\begin{equation}
N = N_0 \exp(-\Gamma_\mathrm{sc} \tau),
\label{eq:fit_eqn}
\end{equation}
where $N_0$ is the initial molecule number and $\tau$ is the illumination time. The photon scattering rate is assumed to be 
$\Gamma_{\mathrm{sc}} =  \Omega_1^2 \Gamma / (1 + 4 ( \Delta / \Gamma )^2 + 2 (\Omega_1 / \Gamma)^2)$, 
where $\Gamma$ is the natural linewidth of the feature and $\Delta$ is the laser detuning from resonance. We normalize Rabi coupling strength to the laser intensity to obtain the reduced transition dipole moment, $\tilde{d}_1 = \Omega_1 / \sqrt{I_0}$. Here, $I_0$ is the peak intensity of the up-leg beam for $\pi$-transitions and half of the peak intensity for $\sigma$ transitions. The linewidths and coupling strengths to the \cS$\ket{v=22, \ J =1}$ hyperfine features are listed in Table~\ref{tab:3_fine_spectroscopy}.

\begin{table}
    \centering
    \caption{ Observed lines for \cSone$\ket{v = 22, J = 1}$ with quantum numbers $m_J$, state labels $m_{I,n}$, measured laser excitation energy $E$, fitted linewidth $\Gamma$, and measured reduced dipole moment $\tilde{d}_1$. The relative uncertainties of the feature locations are 1~MHz. The global uncertainty is 100~MHz, limited by the wavemeter.}
    \begin{tabular}{c  c  c  c  c} 
    \hline \hline
     $m_J$ & $m_{I, n}$ & $E$~(MHz) - 320~THz & $\Gamma / (2 \pi)$~(MHz) & $\tilde{d}_1$~(${\rm kHz} / \sqrt{ \rm mW / cm^2}$)   \\ \hline 
    -1 & $5_0$ & 7611 & 6(2) & 0.24 (3) \\
    -1 & $4_0$ & 7711 & 11(3) & 1.9 (2) \\
    -1 & $4_1$ & 7773 & 5(2) & 1.6 (2) \\ \hline
    
    0 & $5_0$ & 9027 & 4(1) & 0.30(3) \\
    0 & $4_0$ & 9072 & 4(1) & 1.4(2) \\
    0 & $4_1$ & 9102 & 4(1) & 0.42(4) \\
    0 & $3_0$ & 9117 & 7(1) & 4.3(4) \\
    0 & $3_1$ & 9148 & 7(1) & 5.7(6) \\
    0 & $3_2$ & 9176 & 7(1) & 3.0(3) \\ \hline
    
    1 & $3_0$ & 10613 & 5(1) & 2.7(3) \\
    1 & $3_1$ & 10627 & 5(1) & 5.6(6) \\
    1 & $3_2$ & 10643 & 5(1) & 4.4(5) \\
    1 & $4_0$ & 10654 & 5(1) & 2.7(3) \\
    1 & $4_1$ & 10669 & 5(1) & 4.4(5) \\
 
    \hline \hline
    \end{tabular}
    \label{tab:3_fine_spectroscopy}
\end{table}

Which of the observed hyperfine features are suitable for ground state transfer? The Feshbach molecule has a total spin projection of $m_F = 4$. The quantum number that changes in the up-leg transition, depending on the polarization of the STIRAP beam, is $m_F \equiv m_I + m_J$. $\pi$ transitions access $m_F = 4$ excited states, $\sigma^-$ transitions access $m_F = 3$ excited states, and $\sigma^+$ transitions access $m_F = 5$ excited states.  However, at 863.85(1)~G, the excited state is in the Paschen-Back regime and the nuclear spins are decoupled from the electronic spins, as the structure of the observed features show. As a result, it is not possible to change the $m_I$ quantum number with the up-leg laser. We observe strong coupling from the Feshbach state to excited states with $m_{I, n} = 3$ and 4 features and weak couplings to excited states with $m_{I,n}=5$. The Feshbach molecule has dominant closed channel contributions of $m_{I, n} = 3$ and 4, as shown in Table~\ref{tab:1_feshbach_molecules}, which explains the relative coupling strengths to the different hyperfine excited states. In the remainder of this discussion, we refer to states by their $m_I$ labels because $m_I$ is a more constrained quantum number than $m_F$, determining which ground state hyperfine levels can be reached via two-photon transfer. We select \cSone$\ket{v=22, \ J=1, \ m_J=1}$ as the intermediate state for STIRAP as it shows well-resolved features for both $m_{I,n} = 3$ and 4 and a relatively narrow linewidth of $2 \pi \times 5(1)$~MHz.

\section{Transfer to the Ground State}

\begin{figure} 
    \centering
    \includegraphics{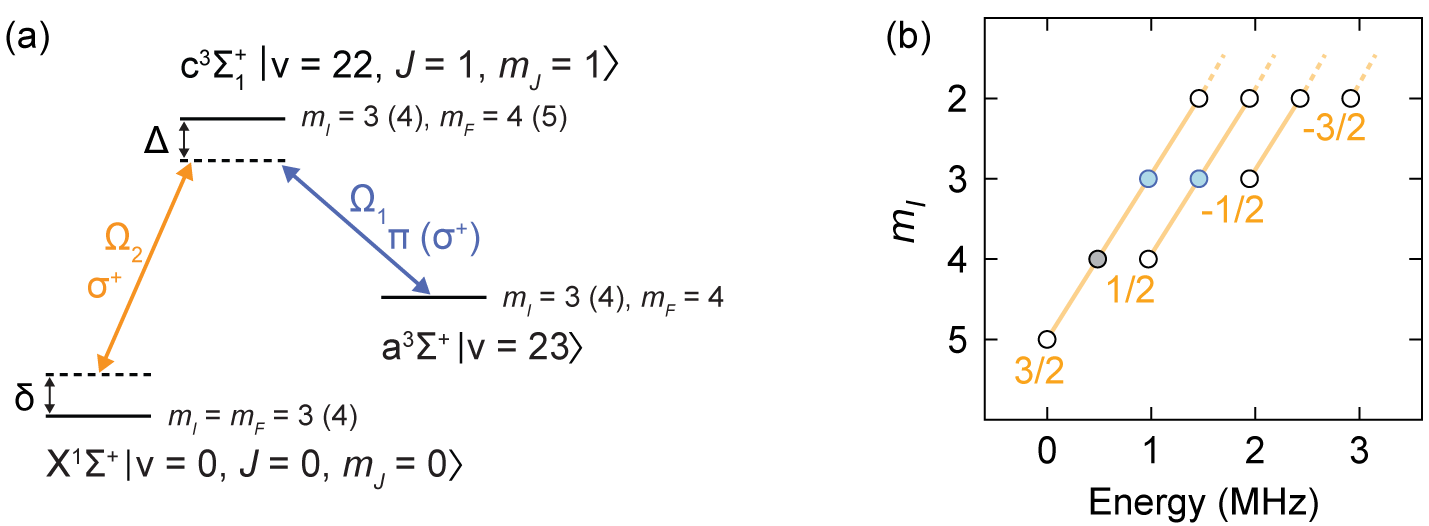}
    \caption{Pathways to different hyperfine ground states of NaCs. (a) Two-photon pathways to ground state levels with $m_I = 3$ and 4. Access to $m_I = 3$ is achieved using a $\pi$-transition on the up-leg and a $\sigma^+$-transition on the down-leg. Access to  $m_I = 4$ is achieved using a $\sigma^+$-transitions on the up-leg and down-leg. $\Delta$ is the one-photon detuning, and $\delta$ the two-photon detuning. (b) Calculated ground state hyperfine structure of NaCs at 864~G. States plotted in blue (grey) are $m_I = m_{I_{\rm Na}} + m_{I_{\rm Cs}} = 3$ (4) levels which are relevant for the STIRAP schemes presented here. Orange diagonal lines connect states with the same $m_{I_{\rm Na}}$ quantum number, labeled in orange.
    }
    \label{fig:6_gs_mF}
\end{figure}

Using $ $\cSone$\ket{v=22, \ J=1, \ m_J=1}$ as the excited state, we perform STIRAP transfer to the rovibrational ground state. In particular, we characterize the transfer efficiency to the ground state levels with $m_I = 3$ and $4$. The corresponding two-photon schemes are illustrated in Figure~\ref{fig:6_gs_mF} (a). The relevant ground state hyperfine structure is shown in Figure~\ref{fig:6_gs_mF} (b). The down-leg laser is a focussed beam (1/$e^2$ radius of 200~$\mu$m) at a wavelength of 642 nm generated by an external-cavity-diode-laser\footnote{ $\Omega_2$ is measured using Autler-Townes spectroscopy.}. 

\begin{figure} 
    \centering
    \includegraphics[width = 3.39 in]{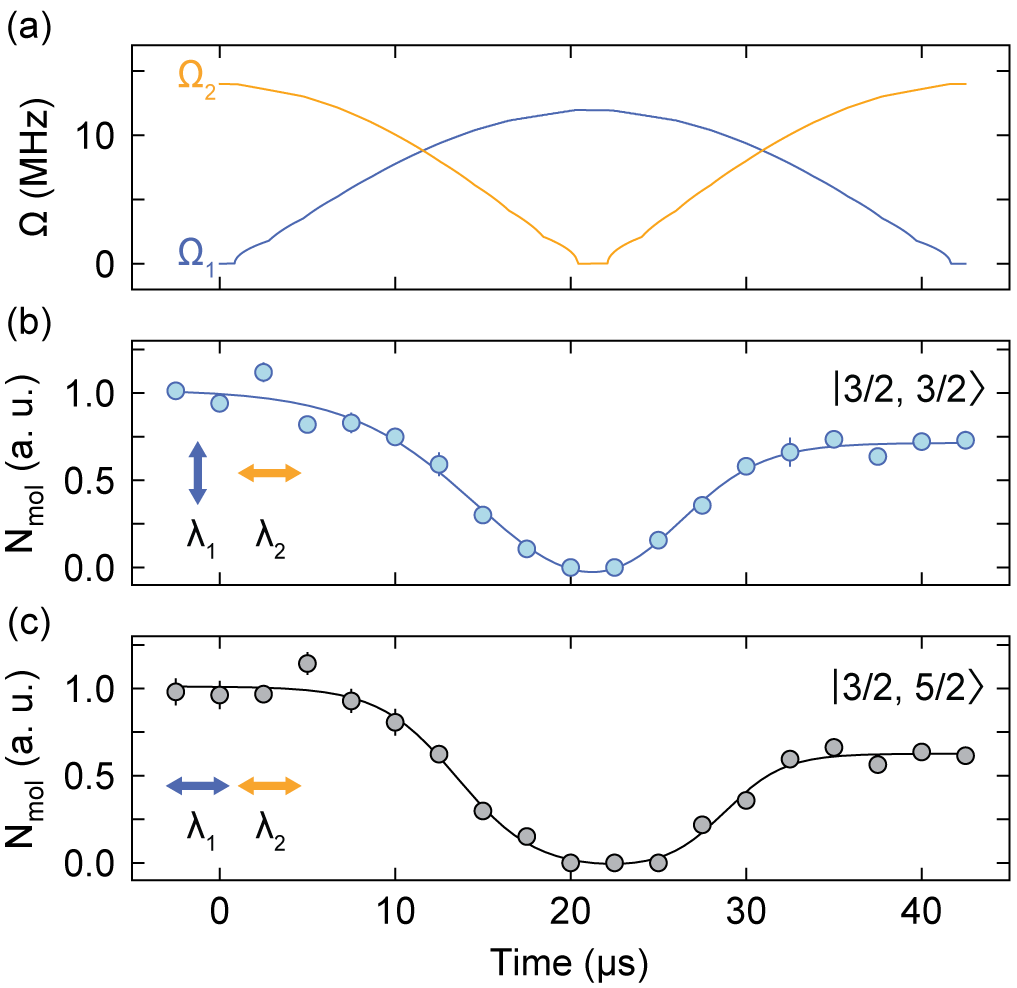}
    \caption{
    STIRAP transfer to and from the ground state. (a) Example of the dynamical change of Rabi frequencies to realize STIRAP transfer. The blue (orange) line indicates the up-leg (down-leg) Rabi frequency, $\Omega_1$ ($\Omega_2$). For transfer to $m_I = 3$, we use $2 \pi \times 12(2)$~MHz for the peak up-leg Rabi frequency, $\Omega_1$, and $2 \pi \times 14(5)$~MHz for the peak down-leg Rabi frequency, $\Omega_2$. For transfer to $m_I = 4$, we use $2 \pi \times 7(1)$~MHz for $\Omega_1$ and $2 \pi \times 14(5)$~MHz for $\Omega_2$. (b) and (c) Feshbach molecule number during the STIRAP sequence using transfer to $\ket{m_{I_{\rm Na}}, \ m_{I_{\rm Cs}}} = \ket{3/2, \ 3/2}$ and $\ket{3/2, \ 5/2}$, respectively. Solid lines are guides to the eye. All data is taken at a one-photon detuning of -180~MHz. 
    }
    \label{fig:7_thereback}
\end{figure}

We apply the pulse sequence shown in Figure~\ref{fig:7_thereback} (a) to the cloud of Feshbach molecules. The first half of the sequence transfers Feshbach molecules to the ground state, and the second half reverses the transfer. We interrupt the pulse sequence at various times and track the population in the Feshbach state, as shown in Figure~\ref{fig:7_thereback} (b) and (c). The fraction of Feshbach molecules that are revived for the fully completed pulse sequence corresponds to the square of the one-way transfer efficiency. Our data corresponds to a one-way transfer efficiency of 88(4)\% for $\ket{m_{I_{\rm Na}}, \ m_{I_{\rm Cs}}} = \ket{3/2, 3/2}$ and 74(3)\% for $ \ket{3/2, 5/2}$, shown here for optimized one-photon detuning, $\Delta$.

\begin{figure} [t]    
    \centering
    \includegraphics[width = 3.39 in]{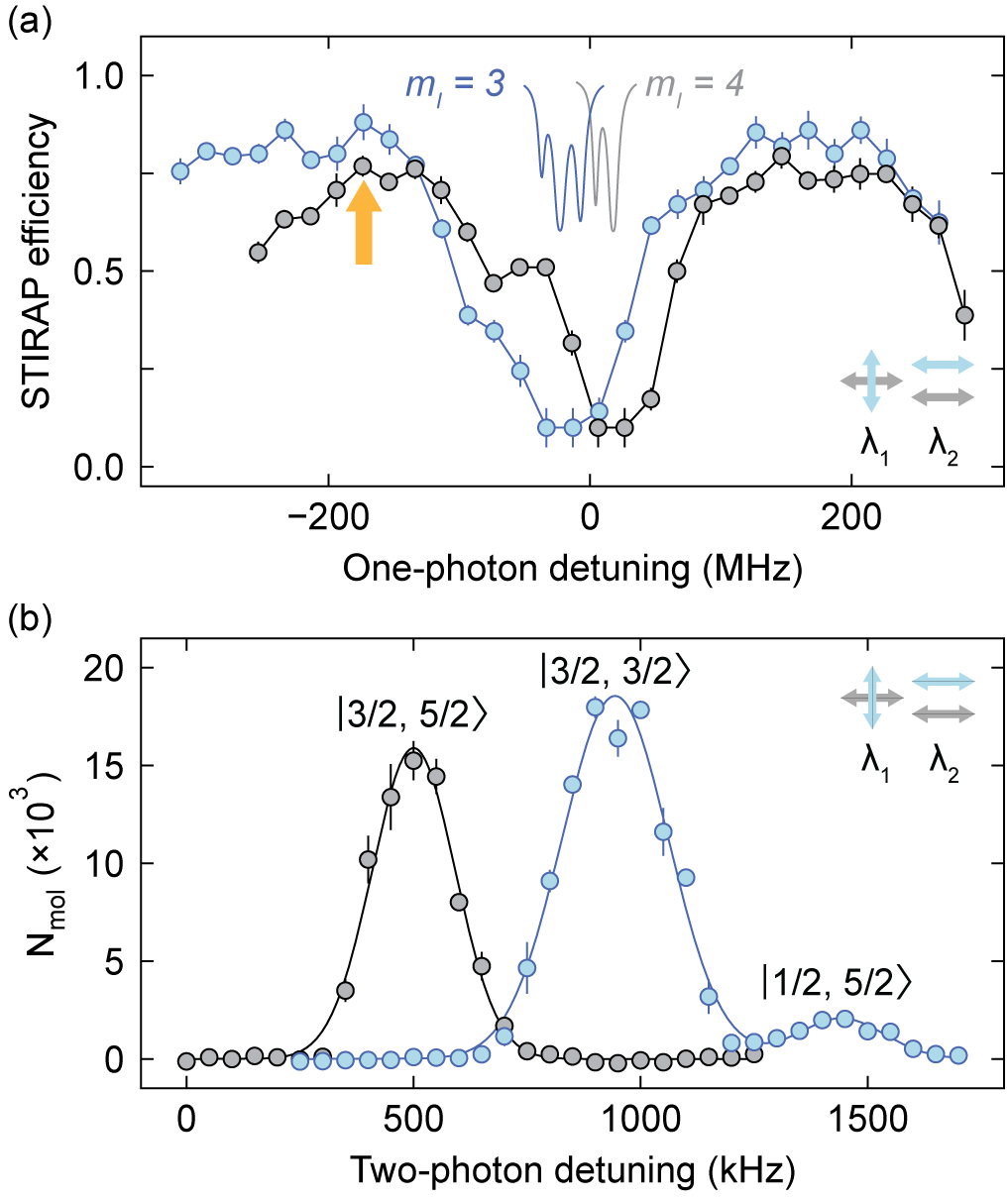}
    \caption{STIRAP efficiency and hyperfine selectivity. (a) Transfer efficiency as a function of one-photon detuning, $\Delta$. Blue (grey) points indicate transfer to the $m_I = 3$ (4) ground state. The $m_I = 3$ (4) excited state structure is overlaid in blue (grey). The orange arrow indicates $\Delta = -180$~MHz, used for the data in (b). The one-photon detuning is measured using the scale in Figure~\ref{fig:5_c22_narrow} (c).  (b) Spectrum of ground state molecules obtained by varying the two-photon detuning, $\delta$, at a fixed $\Delta = -180$~MHz. Blue (grey) points indicate transfer to the $m_I = 3$ (4) ground state. Hyperfine states are labeled with $\ket{ m_{I_{\rm Na}},\ m_{I_{\rm Cs}} }$. The two-photon detuning is measured with respect to the calculated location of the $m_I = 5$ ground state. Curves are guides to the eye. The ground state molecule number is obtained by dividing the Feshbach molecule number by the STIRAP efficiency.
    }
    \label{fig:8_stirap}
\end{figure}

Figure~\ref{fig:8_stirap} (a) shows the transfer efficiency to the ground state as a function of one-photon detuning. Previous work~\cite{cairncross2021assembly, stevenson2022ultracold} has shown that transfer on resonance ($\Delta = 0$) has a very low efficiency for NaCs, which is in contrast to other molecular species, which see optimal transfer on resonance~\cite{ni2008high, park2015ultracold, guo2016creation, molony2014creation, rvachov2017long, bause2021efficient}. Our data confirms that optimal transfer occurs away from resonance, at large positive and negative one-photon detunings, when the detuning is significantly larger than the spacing between the excited state hyperfine features (see Figure~\ref{fig:5_c22_narrow}~(c)). At large detunings, the hyperfine structure is no longer resolved, keeping the system closer to an ideal three-level configuration.

In Figure~\ref{fig:8_stirap} (b), we vary the two-photon detuning and demonstrate access to different ground hyperfine states. For $m_I = 3$, we observe transfer to the $\ket{ m_{I_{\rm Na}},\ m_{I_{\rm Cs}} } = \ket{3/2, \ 3/2}$ and $\ket{1/2, \ 5/2}$ states. For $m_I = 4$, we observe transfer to the $\ket{3/2, \ 5/2}$ state. We observe high transfer efficiencies to the $\ket{3/2, \ 3/2}$ and $\ket{3/2, \ 5/2}$ states, and create samples with more than $1.5 \times 10^4$ ground state molecules. Coupling to the $\ket{1/2, \ 5/2}$ is significantly weaker. We note that the transfer efficiency to the different ground hyperfine states is tied to the nuclear spin composition of the Feshbach state (see Table \ref{tab:1_feshbach_molecules}), with a 31.1\% contribution of $\ket{ S, \ m_S; \ m_{I_{\rm Na}}, \ m_{I_{\rm Cs}}} = \ket{ 1, \ 1; \ 3/2,\  3/2 }$, a 16.8\% contribution of $\ket{ 1, \ 0; \ 3/2,\  5/2 }$, and a 4.9\% contribution of $\ket{ 1, \ 1; \ 1/2,\  5/2 }$. These contributions align with our observation of highly efficient transfer to $\ket{m_{I_{\rm Na}}, \ m_{I_{\rm Cs}}} = \ket{3/2,\ 3/2}$, slightly reduced transfer to $\ket{3/2,\ 5/2}$, and inefficient transfer to $\ket{1/2,\ 5/2}$. The data in Table~\ref{tab:3_fine_spectroscopy} suggests that there should be a very weak, but finite, coupling via the $m_I = 5$ state in the \cS$\ket{v=22, \ J=1, \ m_J=-1}$ excited state manifold. This would be desirable as it would give direct access into the absolute hyperfine ground state, $m_I = 5$. However, we have not observed transfer via that state within the range of accessible experimental parameters.

\section{Conclusion}

We have demonstrated efficient transfer to the $\ket{m_{I_{\rm Na}}, \ m_{I_{\rm Cs}}} = \ket{3/2,\ 3/2}$ and $\ket{3/2,\ 5/2}$ ground states of NaCs via the \cS$\ket{v=22, \ J=1, \ m_J=1}$ intermediate state, improving on previously observed efficiencies. The key to the improved transfer efficiency is an intermediate state with a linewidth of $2\pi \times 5 (1)$~MHz, about 3 times narrower than the intermediate state used in previous work by our group~\cite{stevenson2022ultracold} and nearly 25 times narrower than the intermediate state used for Raman transfer to the ground state for single NaCs molecules~\cite{cairncross2021assembly}. This is aligned with other bialkali molecule studies~\cite{molony2014creation, guo2016creation, bause2021efficient}, which suggest that a narrow excited state linewidth is more important to STIRAP efficiency than strong coupling to the excited state.

In our pursuit of an intermediate state with a narrow linewidth, we benefit from the strong spin-orbit coupling present in NaCs. Avoided crossings in the excited state electronic potentials lead to state mixing in the \cS$\sim \ $\BP complex, providing relatively strong coupling to both the singlet ground state and to the primarily triplet Feshbach state. The avoided crossings also mix the nearly-overlapped \cSone$\ket{v=22}$ and \bP$\ket{v=54}$ states, which narrows the \cSone$\ket{v=22}$ linewidth favorably for STIRAP. In addition to excited states for efficient STIRAP transfer, our spectroscopic survey reveals features with large (>10~GHz) linewidths. We conjecture that these large linewidths may be the result of pre-dissociation, but the precise reason is unknown and merits further investigation. 
 
Improved transfer efficiency to and from the ground state is of central importance for future experiments with many-body quantum systems of NaCs. High transfer efficiency to the ground state allows for the preservation of high phase-space densities of Feshbach molecules~\cite{de2019degenerate}, improving the starting point for direct evaporation to a Bose-Einstein condensate of ground state molecules. In such a system, the superfluid to Mott insulator transition could be used to achieve unit filling of an optical lattice~\cite{greiner2002quantum}. 
 In turn, highly efficient transfer out of the ground state increases detection fidelity. Paired with single-molecule resolved imaging capabilities, this opens the door to cutting-edge studies of ultracold molecules in optical lattices~\cite{capogrosso2010quantum, christakis2023probing} and optical tweezer arrays~\cite{anderegg2019optical, kaufman2021quantum, brooks2021preparation}. 

We thank Eberhard Tiemann for providing a coupled-channel calculation for the Feshbach state and Kang-Kuen Ni and her group for fruitful discussions. We also thank Emily Bellingham for experimental assistance. This work was supported by an NSF CAREER Award (Award No.~1848466), an ONR DURIP Award (Award No.~N00014-21-1-2721) and a Lenfest Junior Faculty Development Grant from Columbia University. C.W.~acknowledges support from the Natural Sciences and Engineering Research Council of Canada (NSERC) and the Chien-Shiung Wu Family Foundation. W.Y.~acknowledges support from the Croucher Foundation. I.S.~was supported by the Ernest Kempton Adams Fund. S.W.~acknowledges additional support from the Alfred P. Sloan Foundation.

\section*{Authors' note}
During the completion of this manuscript, we became aware of complementary work on high-resolution photoassociation spectroscopy of NaCs by the Ni group at Harvard University~\cite{picard2023high}.

\section*{References}
\bibliographystyle{unsrt}

\end{document}